\documentclass[preprintnumbers,showkeys]{revtex4}
\usepackage{amsmath,amsfonts,amssymb,amscd,amsxtra,amsthm}
\usepackage{graphicx}
\usepackage{epstopdf}
\usepackage{bm}
\usepackage{orcidlink}
\usepackage{stackengine}
\usepackage{braket}
\usepackage{ragged2e}
\usepackage{rotating}
\usepackage{booktabs}
\begin{document}
%------------------------------
\preprint{PKNU-NuHaTh-2026}
\title{Flow-Based Global Proposals for Monte Carlo Sampling\\ in SU(2) Lattice Gauge Theory}
%------------------------------
\author{Seung-il Nam\,\orcidlink{0000-0001-9603-9775}}
\email{sinam@pknu.ac.kr}
\altaffiliation{Member of the KLASIQ Collaboration}
\affiliation{Department of Physics, Pukyong National University (PKNU), Busan 48513, Republic of Korea}
%\affiliation{Asia Pacific Center for Theoretical Physics (APCTP), Pohang 37673, Republic of Korea}
%------------------------------
\date{\today}
%------------------------------
\begin{abstract}
We propose a formally valid machine-learning-assisted global proposal mechanism for Monte Carlo sampling in lattice gauge theory. The construction is based on a coupling-flow update on the SU(2) lattice-link manifold, in which active links are transformed conditionally on a frozen-link background. For fixed frozen links, the proposal is explicitly invertible and preserves the product Haar measure, so it can be embedded into a Metropolis-Hastings correction without requiring an explicit model of the full proposal density. We implement the method in two-dimensional pure SU(2) lattice gauge theory and benchmark it against a baseline local Metropolis algorithm used as a controlled reference kernel. In the present testbed, the learned proposal reproduces the target ensemble within statistical resolution across the tested configurations. In matched local-step comparisons, the learned proposal reproduces the target ensemble at a quality comparable to the baseline, but does not outperform the pure local baseline in the conservative matched-step case examined with seed-level statistics within this proof-of-principle setup. At the same time, a favorable mixed-step hybrid configuration yields a modest improvement in effective sample size per unit runtime. Because the learned transformation remains in a near-identity regime, the present results should be interpreted as a proof-of-principle demonstration of formal correctness and limited, configuration-dependent efficiency gain within a controlled comparison, rather than as evidence of superiority over optimized conventional update schemes. This work provides a concrete foundation for extending machine-learned nonlocal updates to larger lattices and non-Abelian gauge theories relevant to lattice QCD.
\end{abstract}
%------------------------------
%------------------------------
\maketitle
%------------------------------
\section{Introduction}
%------------------------------
Markov-chain Monte Carlo (MCMC) sampling based on local updates remains a standard tool in lattice gauge theory, since it provides a nonperturbative route to evaluating path integrals and expectation values of gauge-invariant observables. In practice, however, purely local update schemes suffer from critical slowing down because the Monte Carlo exploration of configuration space proceeds only through small deformations from one configuration to the next. As the continuum limit is approached, the integrated autocorrelation times of observables grow significantly, and for topological observables, this degradation may become particularly severe, leading to the well-known problem of topological freezing~\cite{Kanwar:2024ujc,Bulgarelli:2024cqc,Schaefer:2010hu}. This sampling bottleneck has motivated sustained efforts to design more global, structure-aware update mechanisms.

Recently, machine learning has emerged as a promising tool for accelerating lattice simulations and for constructing nonlocal update proposals that may traverse configuration space more efficiently than conventional local samplers~\cite{Foreman:2023ymy,Kanwar:2024ujc,Cranmer:2023}. In particular, flow-based generative models have attracted considerable attention because they provide invertible transformations with tractable Jacobian factors, making them natural candidates for Monte Carlo proposals and, in favorable cases, for direct approximate samplers of lattice field distributions~\cite{Albergo:2019eim,Albergo:2021vyo,Kanwar:2024ujc}. For gauge theories, this line of development has led to gauge-equivariant and gauge-invariant flow constructions for compact groups, including U(1) and SU($N$), and has recently been extended toward higher-dimensional non-Abelian systems and more scalable architectures~\cite{Kanwar:2020xzo,Boyda:2020hsi,Abbott:2023thq,Abbott:2024knk,Bulgarelli:2024cqc}. These developments suggest that learned nonlocal transformations can help overcome limitations of conventional local samplers.

At the same time, an important distinction must be made between a learned configuration generator and a proposal kernel that can be embedded into a formally correct MCMC algorithm. In a general Metropolis-Hastings (MH) scheme, sampling requires control over both the forward and reverse proposal probabilities so that the acceptance step restores detailed balance with respect to the target distribution~\cite{Tierney1998MH}. Many machine-learning-assisted proposal mechanisms are operationally useful but remain implicit, in the sense that the reverse transition probability is not tractable. In such cases, one may obtain an empirically useful sampler, but not a formally controlled MH update.

In this work, we construct a global proposal mechanism based on an invertible flow transformation acting on SU(2) lattice gauge configurations. The key idea is to design coupling transformations on subsets of lattice links so that the resulting map is explicitly invertible and preserves the product Haar measure on the lattice-link manifold. Because the proposal is then measure preserving and its inverse is known by construction, the MH acceptance probability reduces to the standard Boltzmann weight ratio. This allows us to combine machine-learned nonlocal updates with Monte Carlo sampling, thereby addressing a central limitation of earlier implicit learned proposals~\cite{Albergo:2019eim,Cranmer:2023}. Our formulation is intended as a concrete and practical step toward flow-based Monte Carlo algorithms for non-Abelian lattice gauge theories.

The remainder of this paper is organized as follows. In Sec.~II, we introduce the lattice setup and define the SU(2) gauge theory framework used throughout this work. In Sec.~III, we construct the coupling-flow transformation and establish its invertibility and Haar-measure-preserving properties. In Sec.~IV, we show how this transformation can be embedded into an MH algorithm and derive the corresponding acceptance rule. In Sec.~V, we describe the neural-network parameterization and training procedure used to generate the learned proposal. In Sec.~VI, we present numerical results, including ensemble validation, controlled matched-step comparisons, and a representative mixed-step hybrid benchmark. Finally, Sec.~VII summarizes our findings and discusses possible directions for future work.

%------------------------------
\section{Lattice Setup}
%------------------------------
We consider pure SU(2) lattice gauge theory on a two-dimensional Euclidean square lattice with periodic boundary conditions. The lattice sites are denoted by $x=(x_1,x_2)$, and the lattice directions are labeled by $\mu=1,2$. To each directed link $(x,\mu)$ we assign a gauge variable $U_{x,\mu}\in SU(2)$. The elementary plaquette variable is defined by
\begin{equation}
U_p(x)=U_{x,1}\,U_{x+\hat{1},2}U^\dagger_{x+\hat{2},1}U^\dagger_{x,2},
\end{equation}
where $\hat{\mu}$ denotes the unit vector in the $\mu$ direction. The dynamics is governed by the standard Wilson gauge action~\cite{Wilson:1974},
\begin{equation}
S[U]=\beta \sum_p \left(1-\frac{1}{2}\mathrm{Re}\,\mathrm{Tr}\,U_p\right),
\end{equation}
with inverse coupling $\beta$. The corresponding target probability distribution is
\begin{equation}
\pi(U)=\frac{1}{Z}e^{-S[U]},
\end{equation}
where the partition function is given by
\begin{equation}
Z=\int \prod_{x,\mu} dU_{x,\mu}e^{-S[U]},
\end{equation}
and $dU_{x,\mu}$ denotes the Haar measure on SU(2) for each link.

In the present work, the use of the Haar measure is essential because the correctness of the proposed flow-based update depends on constructing transformations that preserve the product Haar measure on the lattice-link manifold. This allows the learned proposal to be embedded into an MH accept-reject step without introducing an uncontrolled measure factor. As basic gauge-invariant observables, we consider the average plaquette
\begin{equation}
\langle P\rangle =\left\langle\frac{1}{N_p}\sum_p \frac{1}{2}\mathrm{Re}\,\mathrm{Tr}\,U_p\right\rangle,
\end{equation}
where $N_p$ is the total number of plaquettes, together with rectangular Wilson loops
\begin{equation}
W(R,T)=\left\langle\frac{1}{2}\mathrm{Re}\,\mathrm{Tr}\prod_{\ell\in C_{R,T}} U_\ell\right\rangle,
\end{equation}
where $C_{R,T}$ denotes a closed rectangular contour of spatial extent $R$ and temporal extent $T$. In two dimensions, the $1\times1$ Wilson loop coincides with the elementary plaquette, so that $W(1,1)=\langle P\rangle$ by definition. For numerical implementation, each SU(2) link may be represented in quaternion form,
\begin{equation}
U=a_0\mathbf{1}+i\sum_{k=1}^3 a_k \sigma^k,
\end{equation}
with real coefficients satisfying
\begin{equation}
a_0^2+a_1^2+a_2^2+a_3^2=1.
\end{equation}
This representation is convenient for machine-learning applications because the group constraint reduces to a unit-norm condition in $\mathbb{R}^4$, while the group multiplication law remains straightforward to implement numerically.

%------------------------------
\section{Coupling-Flow Construction}
%------------------------------
To construct a machine-learned global proposal, we introduce a coupling-flow transformation on the lattice-link manifold. The essential idea is to update only a selected subset of links while keeping the complementary subset fixed, so that the resulting map is explicitly invertible by construction. Let the full set of lattice links be partitioned into two disjoint subsets,
\begin{equation}
\{U\} = \{U_A\} \cup \{U_B\}, \quad \{U_A\} \cap \{U_B\} = \emptyset.
\end{equation}
Here, $\{U_A\}$ denotes the active subset to be updated, while $\{U_B\}$ denotes the frozen subset that serves as the conditioning background. In practice, such a partition may be chosen according to a checkerboard pattern, an even-odd decomposition, or a more general blockwise scheme. We define a coupling-flow transformation $f_\theta$ by
\begin{equation}
U'_B = U_B, \quad U'_{x,\mu} = G_{x,\mu,\theta}(U_B)\,U_{x,\mu}, \quad (x,\mu)\in A,
\label{eq:flow_AB}
\end{equation}
where $G_{x,\mu,\theta}(U_B)\in SU(2)$ is a learned group element associated with the active link $(x,\mu)$ and determined solely by the frozen subset $U_B$. Equivalently, in collective notation,
\begin{equation}
U'_A = G_\theta(U_B)\,U_A,
\end{equation}
with the understanding that $G_\theta(U_B)$ denotes the set of linkwise SU(2) multipliers acting on the active subset.

The learned multipliers $G_{x,\mu,\theta}$ are generated by a neural network that takes local or multiscale features extracted from the frozen links as input. The crucial structural requirement is that the output on the active subset depends only on $U_B$ and not on $U_A$. This property ensures explicit invertibility of the transformation and is the direct analog of coupling layers in conventional normalizing flows~\cite{Dinh:2017,Papamakarios:2021}. To make the reverse move explicit within the MH construction, we augment the proposal with an auxiliary binary branch variable $s\in\{+1,-1\}$, sampled uniformly at each global proposal. The coupling-flow update is then defined by
\begin{equation}
U'_B = U_B, \quad U'_{x,\mu} = G^{s}_{x,\mu,\theta}(U_B)U_{x,\mu}, \quad (x,\mu)\in A,
\label{eq:branch_update}
\end{equation}
where $G^{+1}_{x,\mu,\theta}(U_B)\equiv G_{x,\mu,\theta}(U_B)$ and
$G^{-1}_{x,\mu,\theta}(U_B)\equiv G_{x,\mu,\theta}(U_B)^{-1}$.
The reverse move is obtained by the opposite branch $-s$, which is selected with the same probability $1/2$. This extended-space construction makes the forward and reverse proposals explicit while preserving the simple coupling-flow form.
%Since the transformation defined in Eq.~(\ref{eq:branch_update}) is invertible and the frozen subset remains unchanged, we have\begin{equation}U_B = U'_B.\end{equation}Hence, the group multipliers $G_{x,\mu,\theta}(U_B)$ are uniquely determined from the transformed configuration through the unchanged background $U_B$. The inverse map on the active subset is therefore\begin{equation}U_{x,\mu}=G^{-s}_{x,\mu,\theta}(U_B)U'_{x,\mu},\quad (x,\mu)\in A,\label{eq:branch_inverse}\end{equation}or, in collective notation,\begin{equation}U_A = G^{-s}_\theta(U_B) U'_A.\end{equation}
Thus, for each fixed branch $s$, the map $f_{\theta,s}:U\mapsto U'$ is bijective. Since the two branches $s=\pm1$ are sampled symmetrically, the proposal on the extended space $(U,s)$ has an explicit inverse and symmetric branch weight.

A second key property of the coupling flow is that it preserves the product Haar measure on the lattice-link manifold. For each active link, the update is given by left multiplication by an SU(2) group element. Since the Haar measure $d\mu(U)$ on SU(2) is left invariant,
\begin{equation}
d\mu(GU)=d\mu(U),\quad G\in SU(2),
\end{equation}
it follows that for every active link, $d\mu(U'_{x,\mu}) = d\mu(U_{x,\mu})$. Because the frozen links are unchanged, their measure is trivially preserved as well. Therefore, the full product measure satisfies
\begin{equation}
\prod_{(x,\mu)\in A} d\mu(U'_{x,\mu})\prod_{(y,\nu)\in B} d\mu(U'_{y,\nu})
=\prod_{(x,\mu)\in A} d\mu(U_{x,\mu})\prod_{(y,\nu)\in B} d\mu(U_{y,\nu}).
\end{equation}
Hence, the coupling transformation preserves the product Haar measure exactly. Because the proposal is both invertible and measure preserving, no additional Jacobian factor arises from the transformation itself~\cite{Papamakarios:2021}. This is the essential structural ingredient behind the MH correction.

%------------------------------
\section{Metropolis-Hastings Acceptance}
%------------------------------
We now show that the coupling-flow proposal can be incorporated into an MH update. In a general MH algorithm, the acceptance probability for a proposal $U\to U'$ is given by
\begin{equation}
A(U\to U')=\min\left[1,\frac{\pi(U')\,q(U|U')}{\pi(U)\,q(U'|U)}\right],
\end{equation}
where $\pi(U)$ is the target distribution and $q(U'|U)$ denotes the proposal kernel. For the present coupling flow, the proposal at fixed partition and fixed branch $s$ is defined by a deterministic bijection $U' = f_{\theta,s}(U)$ whose inverse exists explicitly by construction. Moreover, as shown in the previous section, this map preserves the product Haar reference measure on the lattice-link manifold. In the augmented proposal with branch variable $s$, the forward and reverse branch probabilities are both $1/2$, so they cancel in the Hastings ratio. Therefore, with respect to the Haar reference measure, the forward map and its inverse contribute no Jacobian factor, and the only nontrivial contribution to the acceptance rule is the target weight ratio. In the current implementation, we employ a fixed or periodically cycled checkerboard partitioning scheme to satisfy the required symmetry condition in practice.

Under these conditions, the MH acceptance probability reduces to
\begin{equation}
A(U\to U')=\min\left[1,\frac{\pi(U')}{\pi(U)}\right].
\end{equation}
Using the Wilson gauge weight,
\begin{equation}
\pi(U)\propto e^{-S[U]},
\end{equation}
we obtain the explicit acceptance rule
\begin{equation}
A(U\to U')=\min\left[1,e^{-[S(U')-S(U)]}\right].
\end{equation}
The corresponding transition kernel is therefore
\begin{equation}
K(U\to U') = q(U'|U)A(U\to U'),
\end{equation}
together with the usual rejection probability for staying at $U$. By construction, this kernel satisfies detailed balance,
\begin{equation}
\pi(U)K(U\to U')=\pi(U')K(U'\to U),
\end{equation}
and hence leaves the target distribution invariant. In this sense, the present flow-based proposal defines a formally correct MCMC update despite being generated by a learned nonlocal transformation.

%------------------------------
\section{Neural-Network Parameterization}
%------------------------------
We now specify the concrete implementation used for the learned SU(2) multipliers entering the coupling flow. For each active link $(x,\mu)\in A$, the neural network outputs a three-component real vector,
\begin{equation}
\xi_{x,\mu}^a \in \mathbb{R},\quad a=1,2,3,
\end{equation}
which is interpreted as an element of the Lie algebra $\mathfrak{su}(2)$. The corresponding group element is constructed through the exponential map,
\begin{equation}
G_{x,\mu,\theta}(U_B)=\exp\left[i \sum_{a=1}^3 \xi_{x,\mu}^a(U_B)\sigma^a\right]\in SU(2),
\label{eq:G_from_xi}
\end{equation}
where $\sigma^a$ are the Pauli matrices. In the actual sampler, this multiplier enters the branch-augmented coupling-flow update introduced in Sec.~III,
\begin{equation}
U'_{x,\mu}=G_{x,\mu,\theta}(U_B)^{\,s}\,U_{x,\mu}, \quad (x,\mu)\in A,\quad s\in\{+1,-1\},
\label{eq:branch_update_sec5}
\end{equation}
with the branch variable $s$ sampled uniformly for each global proposal. This parameterization has two immediate advantages. First, the update remains exactly on the SU(2) group manifold by construction, without any need for projection or reunitarization. Second, because the opposite branch corresponds to group inversion, the inverse transformation is obtained simply by
\begin{equation}
U_{x,\mu}=G_{x,\mu,\theta}(U_B)^{-s}\,U'_{x,\mu},
\quad (x,\mu)\in A,
\label{eq:inverse_sec5}
\end{equation}
which is fully consistent with the branch-augmented coupling-flow structure introduced in Sec.~III.

In the present implementation, we employ a compact residual convolutional network with circular padding rather than a graph-based or explicitly gauge-equivariant architecture. The model uses a local feature tensor derived from the staple environment and maps it to the three-component Lie-algebra field $\xi_{x,\mu}^a$. Concretely, the input tensor contains nine channels: four for the forward-plus-backward staple, four for the forward-minus-backward staple, and one for the staple magnitude. The hidden trunk consists of six residual blocks with 128 channels, followed by an output head that produces the three Lie-algebra components on the active checkerboard subset. These outputs are exponentiated to SU(2) multipliers through Eq.~(\ref{eq:G_from_xi}) and then inserted into the branch-augmented update Eq.~(\ref{eq:branch_update_sec5}). The dependence on the frozen subset $U_B$ alone is the essential architectural constraint required for explicit invertibility. More generally, the necessity of the Monte Carlo correction does not depend on whether the underlying architecture is convolutional, graph-based, or gauge equivariant. What is essential is only that the network output defines SU(2) multipliers through the exponential map and that these multipliers depend solely on $U_B$.

For numerical stability, it is convenient to control the magnitude of the Lie-algebra output. We therefore introduce a rescaled parameterization,
\begin{equation}
G_{x,\mu,\theta}(U_B)=\exp\left[i\epsilon\sum_{a=1}^3 \tilde{\xi}_{x,\mu}^a(U_B)\sigma^a\right],
\label{eq:scaled_G}
\end{equation}
where $\epsilon$ is a tunable proposal scale and $\tilde{\xi}_{x,\mu}^a$ denotes the raw network output. This explicit scale parameter allows one to interpolate between conservative proposals with high acceptance and more aggressive proposals with larger collective motion in configuration space. In the branch-augmented sampler, Eq.~(\ref{eq:scaled_G}) is combined with Eq.~(\ref{eq:branch_update_sec5}), so that the two branches correspond to multiplication by $G_{x,\mu,\theta}(U_B)$ and its inverse, respectively. The model is trained in a self-supervised manner using equilibrated configurations generated by the baseline Metropolis sampler. In the production runs reported here, we use Adam with a learning rate of $10^{-3}$, a batch size of 64, 40 epochs, and a 90/10 train-validation split. The training objective is
\begin{equation}
\mathcal{L}=\lambda_\mathrm{action}\left\langle \mathrm{ReLU}(\Delta S)^2\right\rangle
+\lambda_\mathrm{balance}\left\langle \Delta S \right\rangle^2
+\lambda_\mathrm{floor}\mathrm{ReLU}(m_{\mathrm{target}}-m)^2
-\lambda_\mathrm{move}m+\lambda_\mathrm{smooth}\mathcal{L}_\mathrm{smooth},
\end{equation}
with
\begin{equation}
\mathcal{L}_\mathrm{smooth}=\frac{1}{|A|}\sum_{(x,\mu)\in A}\sum_{\hat e}
\left\|\bm{\xi}_{x+\hat e,\mu}-\bm{\xi}_{x,\mu}\right\|^2,
\end{equation}
where $\Delta S$ is the action difference associated with the proposed branch-augmented update and $m$ is the average proposal motion on the active links. Here $\lambda_{\mathrm{action}}$ weights the penalty for positive $\Delta S$, $\lambda_{\mathrm{balance}}$ controls the constraint on the mean action shift, $\lambda_{\mathrm{move}}$ promotes larger proposal distances, $\lambda_{\mathrm{floor}}$ penalizes moves that fall below the target displacement scale, and $\lambda_{\mathrm{smooth}}$ imposes a smoothness regularization on the learned flow. Unless otherwise stated, we use $\lambda_{\mathrm{action}}=1.0$, $\lambda_{\mathrm{balance}}=0.1$, $\lambda_{\mathrm{move}}=0.3$, $\lambda_{\mathrm{floor}}=8.0$, $\lambda_{\mathrm{smooth}}=5.0\times10^{-3}$, and $m_{\mathrm{target}}=1.0\times10^{-2}$, which denotes the target displacement scale.

The checkpoint with the lowest validation loss is used for production runs. The loss suppresses large action increases and drift, encourages finite proposal motion, and regularizes the spatial roughness of the predicted Lie-algebra field. The numerical setup employed in the present study is summarized in Table~\ref{TAB3} in the Appendix. The same lattice size and gauge coupling were used throughout the baseline generation and candidate-sampler evaluation, while the flow-model training was performed on the baseline ensemble with the hyperparameters listed in the middle column.

%------------------------------
\section{Numerical Results}
%------------------------------
We now present numerical tests of the proposed ML-flow-assisted Monte Carlo sampler in two-dimensional pure SU(2) lattice gauge theory. All numerical results reported below are obtained on an $8\times 8$ lattice at $\beta=2.0$. For the baseline ensemble generation, we perform 2000 thermalization sweeps, followed by 2000 saved configurations with 20 sweeps between successive measurements. These simulation parameters are kept fixed throughout training and evaluation unless explicitly stated otherwise. The main goals of the present numerical study are twofold: first, to verify that the learned global proposal preserves the target ensemble within statistical accuracy, and second, to assess whether it can improve sampling efficiency relative to the baseline local Metropolis algorithm. To quantify sampling efficiency, we estimate the integrated autocorrelation time $\tau_{\mathrm{int}}(P)$ of the plaquette and define the effective sample size as
\begin{equation}
\mathrm{ESS}=\frac{N_{\mathrm{cfg}}}{2\tau_{\mathrm{int}}(P)},
\end{equation}
where $N_{\mathrm{cfg}}$ denotes the number of saved configurations. The ESS per second reported below is then obtained by dividing this quantity by the measured wall-clock runtime. In the present study, $\tau_{\mathrm{int}}$ is extracted from the plaquette time series using a standard windowed autocorrelation estimator.

\begin{figure}[t]
\topinset{(a)}{\includegraphics[width=7.5cm]{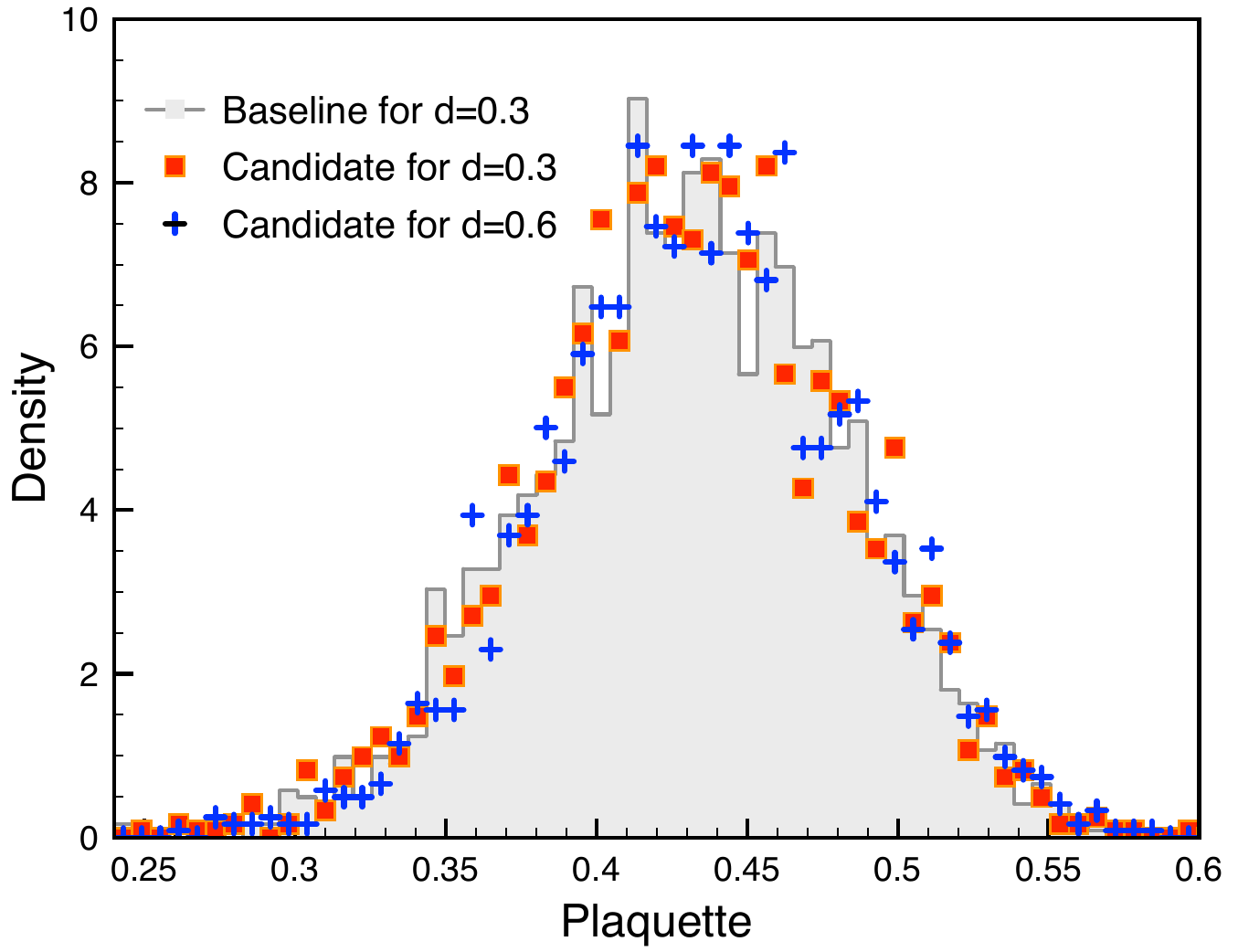}}{-0.4cm}{0.0cm}
\topinset{(b)}{\includegraphics[width=7.5cm]{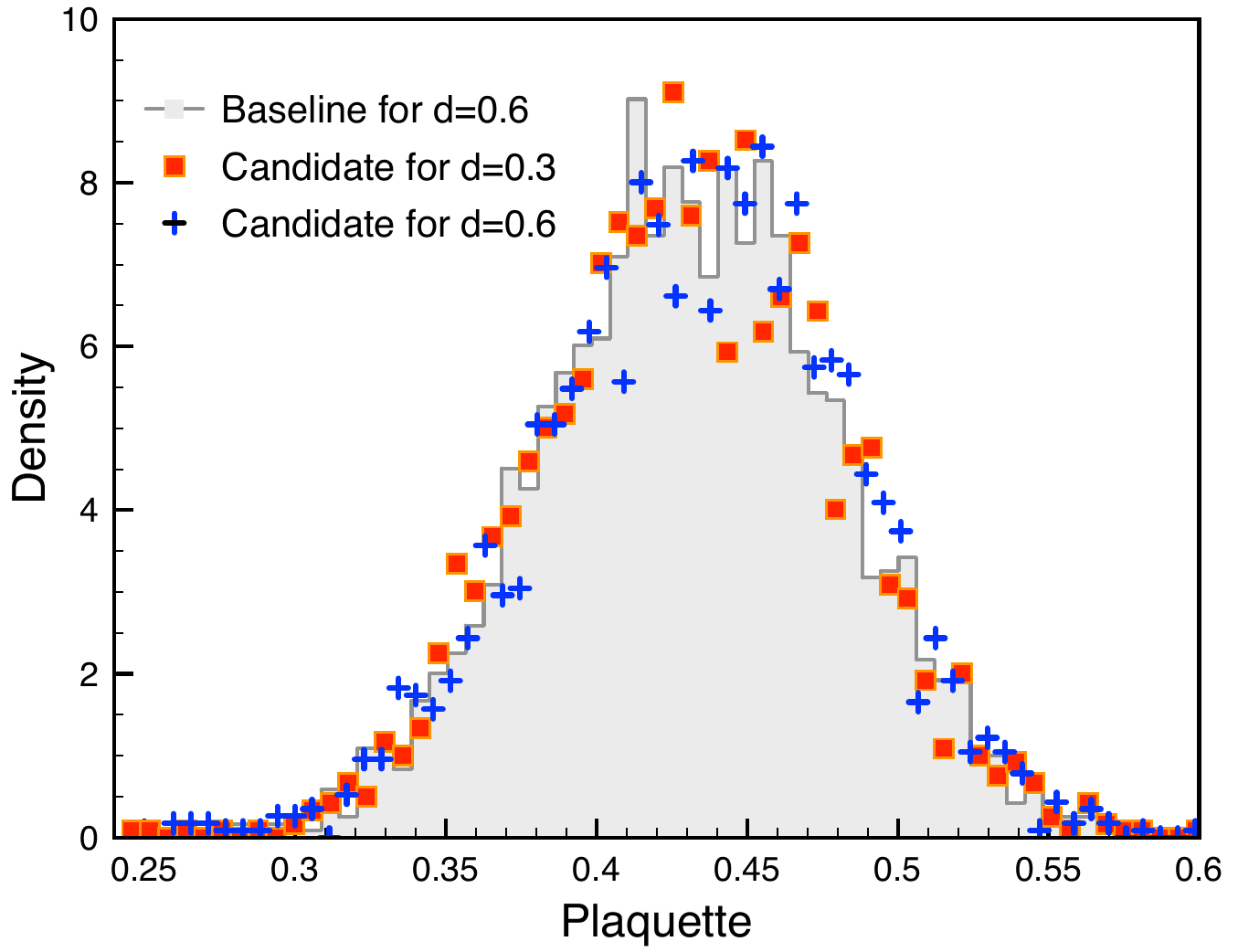}}{-0.4cm}{0.0cm}\\
\vspace{0.5cm}
\topinset{(c)}{\includegraphics[width=7.5cm]{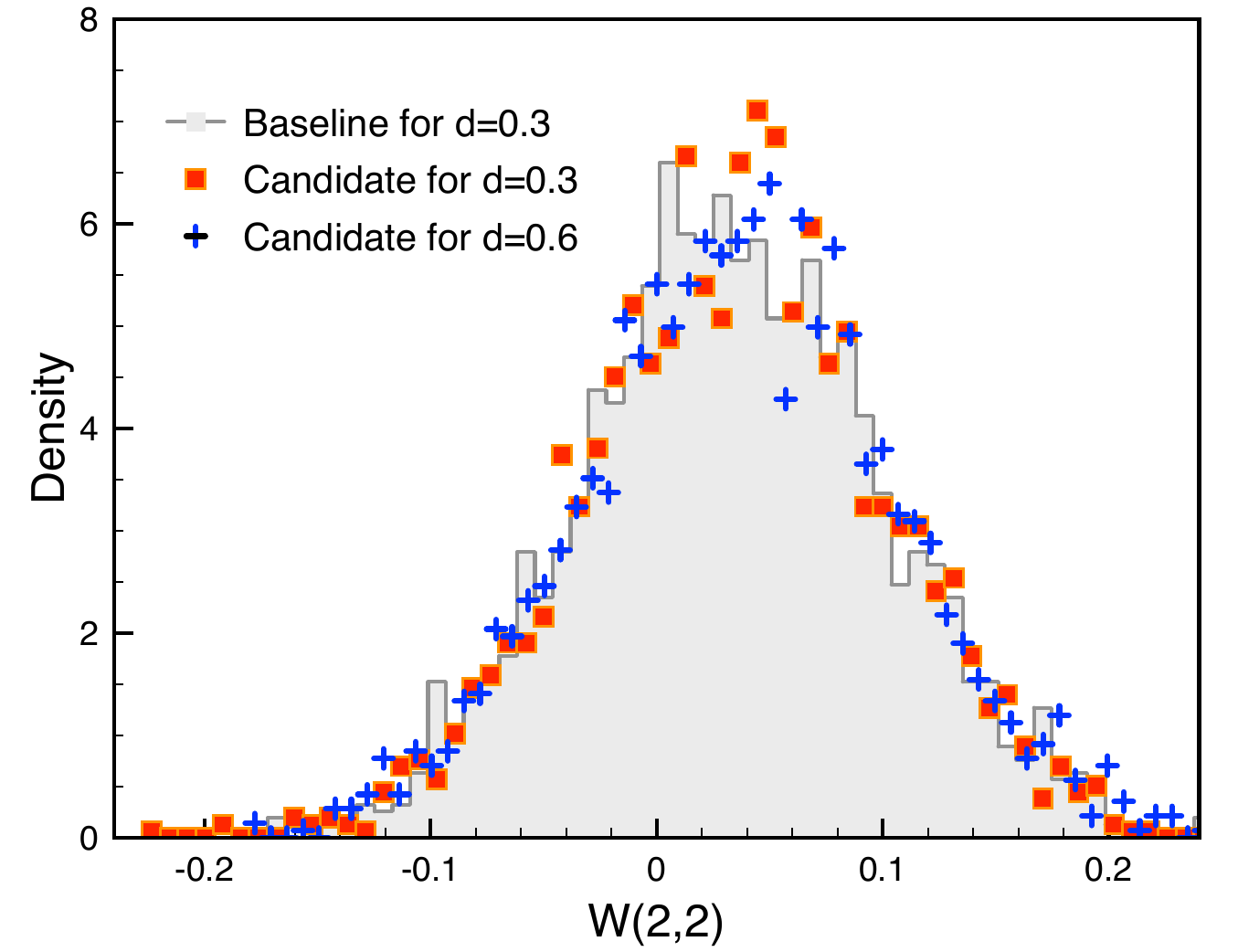}}{-0.4cm}{0.0cm}
\topinset{(d)}{\includegraphics[width=7.5cm]{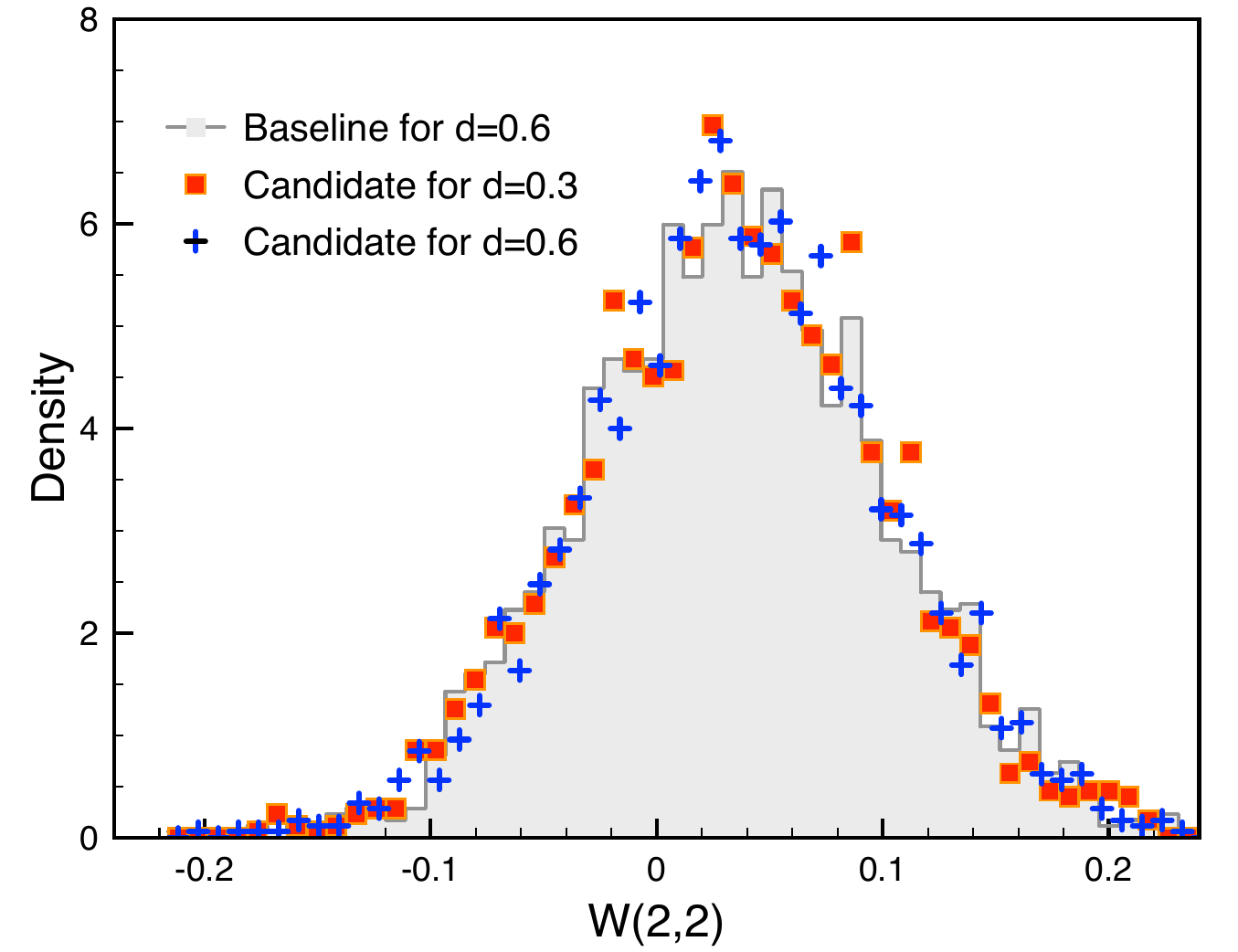}}{-0.4cm}{0.0cm}
\caption{(Color online) Probability density distributions of representative gauge-invariant observables obtained from the baseline local Metropolis (baseline) and the ML-flow-assisted (candidate) samplers in two-dimensional pure SU(2) lattice gauge theory. Panels (a) and (b) show the average plaquette $P$ for baseline local step sizes $d_{\rm b}=0.3$ and $d_{\rm b}=0.6$, respectively, with candidate results for $d_{\rm c}=0.3$ and $d_{\rm c}=0.6$ overlaid in each panel. Panels (c) and (d) show the corresponding Wilson loop $W(2,2)$ for $d_{\rm b}=0.3$ and $d_{\rm b}=0.6$. For the mixed-step case $(d_{\rm b},d_{\rm c})=(0.3,0.6)$ in Table~\ref{TAB}, the corresponding Kolmogorov-Smirnov distances are $\mathrm{KS}(P)=3.00\times10^{-2}$ and $\mathrm{KS}(W_{2,2})=3.35\times10^{-2}$.}
\label{FIG1}
\end{figure}

\begin{figure}[t]
\includegraphics[width=10cm]{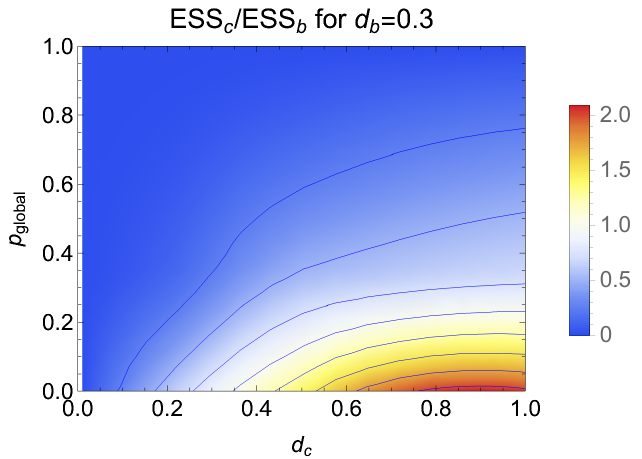}
\caption{(Color online) Density map of $\mathrm{ESS}_c/\mathrm{ESS}_b$ for $d_{\rm b}=0.3$ in the $(d_{\rm c},p_{\mathrm{global}})$ plane.}
\label{FIG3}
\end{figure}

We first examine whether the ML-flow-assisted sampler reproduces the correct distribution of gauge-invariant observables. In Fig.~\ref{FIG1}(a,b), we compare the probability density distributions of the plaquette $P$ obtained from the baseline local Metropolis sampler and from the ML-flow-assisted sampler for two baseline local step sizes, $d_{\rm b}=0.3$ and $d_{\rm b}=0.6$, with both candidate choices $d_{\rm c}=0.3$ and $d_{\rm c}=0.6$ shown for comparison. The corresponding Wilson-loop distributions are shown in Fig.~\ref{FIG1}(c,d). In all four panels, the distributions agree closely within statistical fluctuations. These results indicate that the learned coupling-flow proposal, when combined with the MH correction, produces ensembles consistent with the target distribution. To quantify ensemble agreement, we compute Kolmogorov-Smirnov (KS) distances between the distributions generated by the two samplers. As summarized in Table~\ref{TAB}, the KS distances remain of order $\mathcal{O}(10^{-2})$ across all tested configurations. This shows that, at the level of the observables considered here, the learned proposal does not introduce any statistically significant distortion of the target ensemble. Fig.~\ref{FIG3} shows the efficiency ratio $\mathrm{ESS}_c/\mathrm{ESS}_b$ in the $(d_{\rm c},p_{\mathrm{global}})$ plane for fixed $d_{\rm b}=0.3$. The ratio is larger in the region of relatively large $d_{\rm c}$ and small $p_{\mathrm{global}}$, and decreases as $p_{\mathrm{global}}$ increases. Within the present setup, this indicates that the learned global proposal is most favorable when used infrequently together with a relatively larger candidate local step size. This trend is consistent with the mixed-step result in Table~I, where the configuration $(d_{\rm b},d_{\rm c})=(0.3,0.6)$ gives the largest efficiency ratio.

We next turn to sampling efficiency. Because the candidate sampler mixes a learned global update with a local Metropolis kernel, it is important to distinguish clearly between matched-step comparisons and mixed-step hybrid configurations. In the present work, Table~\ref{TAB} serves as the main controlled comparison across the four combinations of baseline local step size $d_{\rm b}$ and candidate local step size $d_{\rm c}$. The fairest like-for-like comparisons are the matched-step cases $(d_{\rm b},d_{\rm c})=(0.3,0.3)$ and $(0.6,0.6)$. In these cases, the candidate ESS/sec is comparable to, but not larger than, the corresponding baseline value, with ratios $0.916$ and $0.951$, respectively. Thus, within the present small testbed, the matched-step comparison supports formal viability and comparable performance rather than a decisive efficiency gain. By contrast, the mixed-step hybrid configurations demonstrate that the overall efficiency can depend strongly on the interplay between local-step tuning and the learned global proposal. In particular, the configuration $(d_{\rm b},d_{\rm c})=(0.3,0.6)$ yields ${\rm ESS}_{\rm candidate}/{\rm ESS}_{\rm baseline}=1.728$, whereas $(d_{\rm b},d_{\rm c})=(0.6,0.3)$ yields ${\rm ESS}_{\rm candidate}/{\rm ESS}_{\rm baseline}=0.506$. These results show that sampling efficiency can benefit from a favorable hybrid configuration involving the learned global proposal, but they also make clear that the observed gain reflects the full kernel composition rather than the learned proposal acting in isolation.
\begin{table}[b]
\centering
\setlength{\tabcolsep}{7pt}
\begin{tabular}{lcccc}
\hline\hline
Baseline local step size $d_{\rm b}$ & $0.3$ & $0.3$ & $0.6$ & $0.6$ \\
Candidate local step size $d_{\rm c}$ & $0.3$ & $0.6$ & $0.3$ & $0.6$ \\
Number of flow cycles & $2$ & $2$ & $2$ & $2$ \\
Global proposal probability $p_{\mathrm{global}}$ & $0.01$ & $0.01$ & $0.01$ & $0.01$ \\
Proposal scale & $0.6$ & $0.6$ & $0.6$ & $0.6$ \\
\hline
${\rm ESS}_{\rm baseline}$/sec & $5.849$ & $5.849$ & $11.230$ & $11.230$ \\
${\rm ESS}_{\rm candidate}$/sec & $5.360$ & $10.106$ & $5.685$ & $10.676$ \\
${\rm ESS}_{\rm candidate}/{\rm ESS}_{\rm baseline}$ & $\textbf{0.916}$ & $\textbf{1.728}$ & $\textbf{0.506}$ & $\textbf{0.951}$ \\
$\tau_{\rm int}[P]_{\rm baseline}$ & $1.010$ & $1.010$ & $0.538$ & $0.538$ \\
$\tau_{\rm int}[P]_{\rm candidate}$ & $1.022$ & $0.543$ & $0.973$ & $0.512$ \\
\hline
Global move average & $1.72\times10^{-6}$ & $1.73\times10^{-6}$ & $3.77\times10^{-7}$ & $3.68\times10^{-7}$ \\
Active move average & $4.95\times10^{-7}$ & $4.99\times10^{-7}$ & $1.02\times10^{-7}$ & $9.93\times10^{-8}$ \\
Branch mean & $0.039$ & $-0.031$ & $-0.035$ & $-0.030$ \\
\hline
$\langle P\rangle$ & $4.308\times10^{-1}$ & $4.341\times10^{-1}$ & $4.306\times10^{-1}$ & $4.328\times10^{-1}$ \\
$\langle W_{22}\rangle$ & $3.421\times10^{-2}$ & $3.643\times10^{-2}$ & $3.415\times10^{-2}$ & $3.572\times10^{-2}$ \\
KS$(P)$ & $2.250\times10^{-2}$ & $3.000\times10^{-2}$ & $1.550\times10^{-2}$ & $3.250\times10^{-2}$ \\
KS$(W_{22})$ & $3.000\times10^{-2}$ & $3.350\times10^{-2}$ & $1.700\times10^{-2}$ & $1.850\times10^{-2}$ \\
\hline\hline
\end{tabular}
\caption{Unified comparison of the four hybrid-update configurations $(d_{\rm b},d_{\rm c})=(0.3,0.3)$, $(0.3,0.6)$, $(0.6,0.3)$, and $(0.6,0.6)$. Here $d_{\rm b}$ and $d_{\rm c}$ denote the baseline and candidate local Metropolis step sizes, respectively. The matched-step cases $(0.3,0.3)$ and $(0.6,0.6)$ provide the fairest like-for-like reference points, while the mixed-step cases illustrate how the overall efficiency depends on the full hybrid kernel configuration. The diagnostics \textit{Global move}, \textit{Active move}, and \textit{Branch mean} are defined in Appendix~\ref{app:diagnostics}. The branch mean remains close to zero in all cases, indicating approximately balanced sampling of the auxiliary branch variable $s\in\{+1,-1\}$.}
\label{TAB}
\end{table}

Among the mixed-step cases in Table~\ref{TAB}, the configuration $(d_{\rm b},d_{\rm c})=(0.3,0.6)$ yields the largest efficiency ratio. For this case, the candidate ESS/sec is $10.106$ compared with the baseline value $5.849$, corresponding to ${\rm ESS}_{\rm candidate}/{\rm ESS}_{\rm baseline}=1.728$. We emphasize that this number should be read as a performance result for a favorable mixed-step hybrid configuration, not as a like-for-like matched-step comparison and not as a clean measure of the isolated ML effect. The same mixed-step case also illustrates the near-identity character of the learned proposal. In Table~\ref{TAB}, the mean action variation remains modest, as indicated by the small KS distances and the small global-move diagnostic, while the resulting sampler nonetheless exhibits reduced autocorrelation compared with the corresponding baseline configuration. Accordingly, the observed efficiency gain should be interpreted with caution. At this stage, we cannot cleanly separate the contribution of genuinely structured nonlocal deformation from that of proposal scheduling and hybrid-kernel mixing, because no dedicated ablation study has yet been performed. Therefore, the present results establish that the learned proposal can be incorporated consistently into the Markov chain and can coincide with improved decorrelation in this testbed, but they do not yet isolate the precise mechanism responsible for the observed gain.

\begin{figure}[t]
\topinset{(a)}{\includegraphics[width=7.5cm]{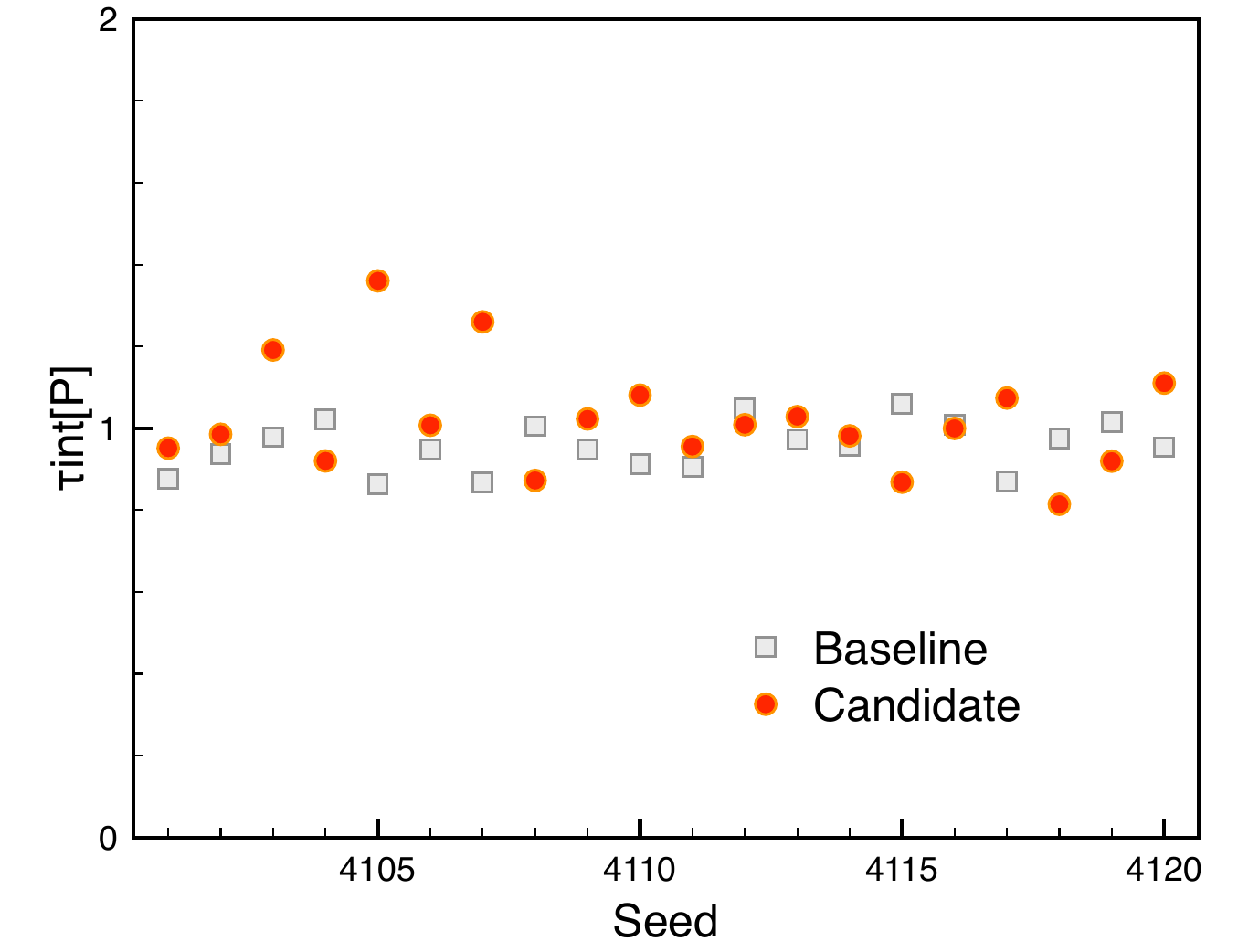}}{-0.4cm}{0.0cm}
\topinset{(b)}{\includegraphics[width=7.5cm]{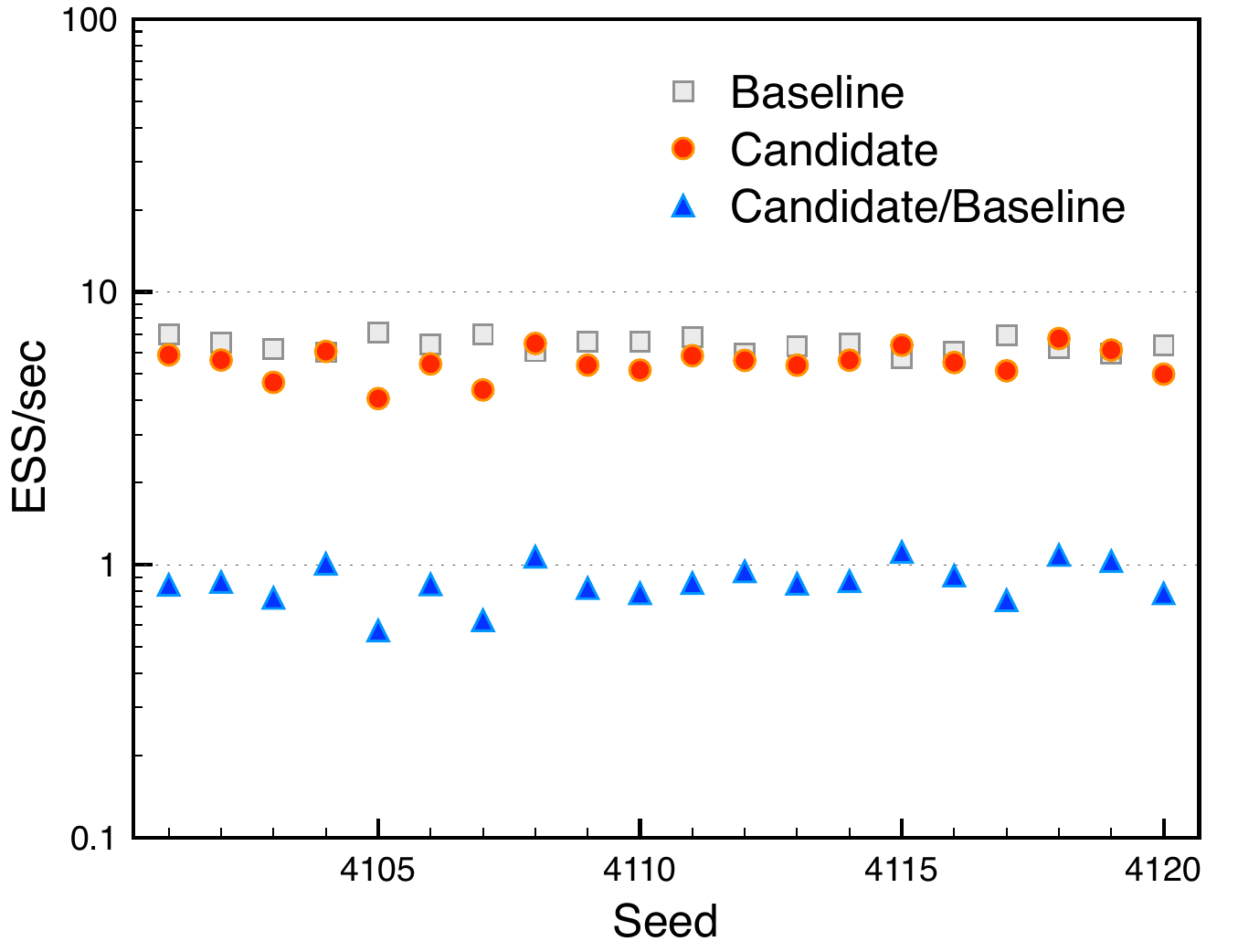}}{-0.4cm}{0.0cm}
\caption{(Color online) Seed-by-seed variation of (a) the integrated autocorrelation time $\tau_{\mathrm{int}}[P]$ and (b) the plaquette ESS/sec for the local Metropolis (baseline) and the ML-flow-assisted (candidate) samplers in the matched-step configuration $(d_{\rm b},d_{\rm c})=(0.3,0.3)$. Panel (b) also shows the seed-by-seed ratio ${\rm ESS}_{\rm candidate}/{\rm ESS}_{\rm baseline}$.}
\label{FIG2}
\end{figure}

We emphasize that the small numerical value of the reported global-move diagnostic should not be misinterpreted as evidence that the learned proposal is negligible. In our implementation, the quantity listed as the global move in Table~\ref{TAB} is the mean-squared displacement between the initial and final configurations, averaged over all lattice sites, both link directions, and all quaternion components. For the $8\times8$ lattice, this corresponds to averaging over $8\times8\times2\times4=512$ degrees of freedom, so the resulting number is naturally small. Accordingly, it is expected to be smaller than the final global displacement even when the proposal is nontrivial.  To assess robustness under a like-for-like comparison, we also examine seed-by-seed variations for the matched-step configuration $(d_{\rm b},d_{\rm c})=(0.3,0.3)$. The results are shown in Fig.~\ref{FIG2}, where we compare the integrated autocorrelation time $\tau_{\mathrm{int}}(P)$ across multiple independent random seeds for both the baseline local Metropolis sampler and the ML-flow-assisted sampler under the same local step size. As seen in Fig.~\ref{FIG2}, the candidate results are broadly comparable to the baseline results, with visible seed-to-seed fluctuations and no statistically significant systematic advantage. Figure~\ref{FIG2} should therefore be read as a robustness diagnostic for the matched-step comparison, consistent with the proof-of-principle character of the present study.

As shown in Table~\ref{TAB4}, for the reversible-branch sampler with $(d_{\rm b},d_{\rm c})=(0.3,0.3)$, the seed-averaged autocorrelation time is slightly larger for the candidate than for the baseline. Over 20 independent seeds, we find
\begin{equation}
\tau_{\mathrm{int}}[P]_{\rm candidate}=1.0206 \pm 0.1332,
\quad
\tau_{\mathrm{int}}[P]_{\rm baseline}=0.9562 \pm 0.0600,
\end{equation}
with the corresponding 95\% bootstrap confidence intervals shown in Table~\ref{TAB4}. Consistent with this trend, the seed-averaged efficiency ratio remains below unity,
\begin{equation}
\mathrm{ESS}_{\mathrm{candidate}}/\mathrm{ESS}_{\mathrm{baseline}}=0.8688 \pm 0.1438,
\end{equation}
with a 95\% bootstrap confidence interval of $[0.8074,\,0.9287]$. This indicates that, for this matched-step choice, the present candidate kernel remains statistically less efficient than the pure local baseline, although the gap is noticeably smaller than the $0.916$ ratio reported in Table~\ref{TAB} for the same configuration. 

\begin{table}[b]
\begin{tabular}{lcc}
\hline\hline
Metric & Mean $\pm$ seed s.d. & 95\% bootstrap CI \\
\hline
$\tau_{\mathrm{int}}[P]_{\rm candidate}$ & $1.0206 \pm 0.1332$ & $[0.9672,\,1.0800]$ \\
$\tau_{\mathrm{int}}[P]_{\rm baseline}$ & $0.9562 \pm 0.0600$ & $[0.9306,\,0.9811]$ \\
\hline\hline
\end{tabular}
\caption{Seed-level uncertainty summary for the integrated autocorrelation time in the matched-step case $(d_{\rm b},d_{\rm c})=(0.3,0.3)$.}
\label{TAB4}
\end{table}

Taken together, the present results demonstrate that the proposed coupling-flow-based global update can be consistently embedded into an MH framework while preserving the target ensemble. The matched-step comparisons show that the learned proposal remains comparable in quality to the baseline without degrading ensemble agreement, although the seed-level statistics for $(d_{\rm b},d_{\rm c})=(0.3,0.3)$ indicate that the candidate is not more efficient than the pure local baseline in that conservative setting. The mixed-step hybrid cases further show that, within a favorable configuration of local and global updates, the ESS per second can improve relative to the chosen baseline reference run while maintaining small KS distances for both the plaquette and the Wilson loop. However, this numerical gain should be interpreted conservatively. Because the learned proposal remains in a near-identity regime and because the current study does not include ablations with matched proposal frequency or alternative hybrid schedules, the origin of the gain cannot yet be uniquely attributed to large-scale nonlocal transport in configuration space. Overall, the present proof-of-principle study shows that machine-learned coupling flows provide a viable and formally controlled route toward learned global proposals in non-Abelian lattice gauge theory, while also making clear that stronger efficiency claims will require broader comparisons and more diagnostic analysis.

%------------------------------
\section{Conclusion}
%------------------------------
In this work, we have developed a machine-learning-assisted global proposal mechanism for Monte Carlo sampling in non-Abelian lattice gauge theory. The key idea is to formulate a coupling-flow transformation on the SU(2) lattice-link manifold, in which a subset of links is updated conditionally on a frozen background. By design, the resulting map is explicitly invertible and preserves the product Haar measure, allowing it to be combined with a Metropolis-Hastings accept-reject step without requiring explicit evaluation of the proposal density. In this sense, the main technical contribution of the present work is the formal construction of a learned proposal that can be embedded in a provably correct MH update on a non-Abelian lattice gauge manifold without explicitly modeling the full proposal density. We have implemented the method in two-dimensional pure SU(2) lattice gauge theory and performed a systematic comparison with the conventional local Metropolis sampler. The numerical results demonstrate that the ML-flow-assisted sampler reproduces the target ensemble within statistical accuracy, as verified by agreement between the plaquette and Wilson-loop distributions and by Kolmogorov-Smirnov diagnostics. In matched-step comparisons, the learned proposal reproduces the target ensemble at a quality comparable to the baseline within the present proof-of-principle setup, but it does not outperform the pure local baseline in the conservative matched-step case examined with seed-level statistics. At the same time, selected mixed-step hybrid configurations show that modest efficiency gains are possible within a favorable choice of local and global update parameters.

A detailed analysis of the proposal diagnostics shows that the learned transformation operates in a near-identity regime, characterized by a very high acceptance rate and small action variations. This conservative behavior is encouraging from the standpoint of formal correctness and stability, but it also limits how strongly the present efficiency gains can be interpreted. In particular, the current results do not yet isolate whether the observed improvement in the favorable mixed-step case is driven primarily by learned structured deformation, by proposal scheduling, or by the hybrid composition of local and global moves. The present study should therefore be viewed as a proof of principle that machine-learned coupling flows can be integrated into a formally correct Monte Carlo framework for non-Abelian gauge theories. Unlike many existing learned proposal mechanisms, the present construction explicitly maintains invertibility and measure preservation, thereby ensuring formal correctness while retaining the flexibility of data-driven updates. Several directions for future work are suggested by the present results. A limitation of the present study is that the ML-assisted sampler is compared only with the standard local Metropolis algorithm. At this stage, we do not claim superiority over optimized SU(2) heatbath or overrelaxation updates. Rather, the purpose here is to establish a controlled proof-of-principle comparison against the baseline kernel used both for ensemble generation and for the reference MCMC dynamics. This framing is essential for correctly interpreting the current performance numbers: they quantify improvement relative to the chosen reference kernel, not relative to the state of the art. A systematic comparison with heatbath- and overrelaxation-based update schemes is therefore an important next step.

First, increasing the effective proposal magnitude while maintaining a reasonable acceptance rate may yield larger gains in sampling efficiency. This may be achieved through improved network architectures, multiscale feature representations, or more expressive coupling structures. Second, the current training objective contains several competing terms, and the present study does not yet provide a systematic sensitivity analysis of the associated hyperparameters. A useful extension would therefore be to report the dependence of acceptance, move size, and ESS on the coefficients entering the loss function, so that the balance between stability and proposal strength can be understood more transparently. Third, extending the present framework to higher-dimensional lattices and SU(3) gauge theory will be essential for assessing its practical impact in realistic lattice QCD simulations~\cite{Abbott:2024corr}. In particular, the present $8\times 8$ two-dimensional testbed is too small and too benign to probe the regimes in which critical slowing down and topological freezing become truly severe. On larger lattices and closer to the continuum limit, the principal open questions will concern the scaling of training cost, the maintenance of acceptance under larger collective moves, and the preservation of useful proposal magnitude as the physical correlation length grows. A systematic comparison with more advanced update schemes, such as heatbath, overrelaxation, trivializing gradient flows~\cite{Bacchio:2023}, or mixed local-update strategies, is also left for future work.

Finally, incorporating irreversible or lifted dynamics into the flow-based proposal may provide a route toward further reducing autocorrelation times beyond the limits imposed by detailed balance. Overall, the present results establish that machine-learned coupling flows provide a formally controlled and practically viable starting point for constructing learned global proposals in non-Abelian lattice gauge theory. The numerical gains observed here remain modest, configuration-dependent, and limited to a small proof-of-principle testbed, but the formal construction developed in this work provides a concrete foundation for future extensions to larger lattices and to more physically demanding regimes, such as lattice QCD.

%------------------------------
\section*{Acknowledgment}
%------------------------------
The author is grateful to the members of the KLASIQ Collaboration for valuable discussions, in particular Jong-Wan Lee (CTPU/IBS), Jangho Kim (Seoul National University), and Alexander Rothkopf (Korea University). This work was supported by the National Research Foundation of Korea (NRF), funded by the Korean government (MSIT), under Grant No.~RS-2025-16065906.

%------------------------------
\appendix
%------------------------------

%------------------------------
\section{Update Process of the ML-Flow+MH Sampler}
%------------------------------
For clarity, the ML-flow-assisted Monte Carlo update may be summarized as follows:
\begin{itemize}
\item \textbf{Initialization:} Begin from a gauge configuration $U$.
\item \textbf{Partitioning:} Decompose the lattice links into an active subset $A$ and a frozen subset $B$.
\item \textbf{Branch selection:} Sample an auxiliary branch variable $s\in\{+1,-1\}$ uniformly.
\item \textbf{Network evaluation:} Using only the frozen subset $U_B$, evaluate the neural network and construct the SU(2) multipliers $G_{x,\mu,\theta}(U_B)$ for all active links.
\item \textbf{Proposal construction:} Build the candidate configuration according to
\begin{align}
U'_B &= U_B, \nonumber\\
U'_{x,\mu} &= G_{x,\mu,\theta}(U_B)^s\,U_{x,\mu}, \quad (x,\mu)\in A.
\end{align}
\item \textbf{Acceptance test:} Compute the action difference
\begin{equation}
\Delta S = S(U') - S(U),
\end{equation}
and accept the proposal with probability
\begin{equation}
A(U\to U') = \min\left(1, e^{-\Delta S}\right).
\end{equation}
If the proposal is rejected, retain the original configuration $U$.
\item \textbf{Iteration:} Change the partitioning pattern and repeat the update so that all links are covered across successive coupling-flow steps.
\end{itemize}

%------------------------------
\section{Diagnostic Quantities}
\label{app:diagnostics}
%------------------------------
For clarity, we collect here the diagnostic quantities quoted in Tables~\ref{TAB} and \ref{TAB4}. The \textit{global move} is defined as the mean-squared displacement between the initial and final lattice configurations,
\begin{equation}
M_{\rm global}=\frac{1}{L_xL_y\,N_\mu\,N_q}\sum_{x,\mu,a}\left(U'_{x,\mu,a}-U_{x,\mu,a}\right)^2,
\end{equation}
where $N_\mu=2$ is the number of link directions and $N_q=4$ is the number of quaternion components per link.

The \textit{active move} is the same quantity restricted to the active subset of links,
\begin{equation}
M_{\rm active}=\frac{1}{|A|\,N_q}\sum_{(x,\mu)\in A}\sum_a\left(U'_{x,\mu,a}-U_{x,\mu,a}\right)^2.
\end{equation}

Finally, the \textit{branch mean} is the sample mean of the auxiliary binary branch variable,
\begin{equation}
\langle s\rangle=\frac{1}{N_{\rm prop}}\sum_{i=1}^{N_{\rm prop}} s_i,
\quad s_i\in\{+1,-1\},
\end{equation}
which should remain close to zero when the two branches are sampled in a balanced way. Unless otherwise stated, these displacements are evaluated on the proposed configuration pair $(U, U')$ prior to the Metropolis accept-reject decision and are then averaged over proposal attempts.

%------------------------------
\section{Loss-Function Coefficients and Reproducibility}
%------------------------------
The training loss used in the present work combines several terms that play distinct roles: suppression of energetically unfavorable proposals, reduction of systematic drift, maintenance of nonzero proposal magnitude, and spatial regularization of the predicted Lie-algebra field. The coefficients adopted in the production run were selected empirically to achieve stable training together with high acceptance in the downstream MH-corrected sampler. We do not claim that these values are optimal. A full sensitivity analysis is beyond the scope of the present proof-of-principle study, and this limitation should be explicitly stated. In particular, the relative weights $\lambda_{\mathrm{action}}$, $\lambda_{\mathrm{balance}}$, $\lambda_{\mathrm{move}}$, $\lambda_{\mathrm{floor}}$, and $\lambda_{\mathrm{smooth}}$ can affect the trade-off among acceptance, move size, and ESS, and a broader scan over these coefficients will be necessary in future work to clarify robustness and reproducibility.

%------------------------------
\section{Numerical Setup}
%------------------------------
The numerical setup used in the present study is summarized in Table~\ref{TAB3}. The same lattice size and gauge coupling were used throughout the baseline generation, flow-model training, and candidate-sampler evaluation so that the impact of the learned global proposal could be assessed under controlled conditions. The baseline and candidate local Metropolis step sizes are varied separately in the main comparison table, Table~\ref{TAB}, and are therefore not listed again here. All wall-clock timings entering ESS/sec were measured in the same software and hardware environment, using the same runtime accounting convention for both baseline and candidate samplers.

\begin{sidewaystable}
\centering
\setlength{\tabcolsep}{18pt}
\begin{tabular}{lccc}
\hline\hline
Parameter & Local Metropolis (Baseline) & ML training & ML-flow-assisted (Candidate) \\
\hline
Lattice size & $8\times8$ & --- & $8\times8$ \\
Gauge coupling & $\beta = 2.0$ & --- & $\beta = 2.0$ \\
Training data & --- & Baseline & --- \\
Thermalization sweeps & $2000$ & --- & $2000$ \\
Saved configurations & $2000$ & --- & $2000$ \\
Sweeps between measurements & $20$ & --- & $20$ \\
Batch size & --- & $64$ & --- \\
Epochs & --- & $40$ & --- \\
Learning rate & --- & $1.0\times10^{-3}$ & --- \\
Hidden channels & --- & $128$ & --- \\
Residual blocks & --- & $6$ & --- \\
Proposal scale & --- & $0.6$ & $0.6$ \\
Number of flow cycles & --- & $2$ & $2$ \\
$\lambda_{\mathrm{action}}$ & --- & $1.0$ & --- \\
$\lambda_{\mathrm{balance}}$ & --- & $0.1$ & --- \\
$\lambda_{\mathrm{move}}$ & --- & $0.3$ & --- \\
$\lambda_{\mathrm{floor}}$ & --- & $8.0$ & --- \\
Move target & --- & $1.0\times10^{-2}$ & --- \\
$\lambda_{\mathrm{smooth}}$ & --- & $5.0\times10^{-3}$ & --- \\
Global proposal probability $p_{\mathrm{global}}$ & --- & --- & $1.0\times10^{-2}$ \\
\hline\hline
\end{tabular}
\caption{Numerical setup common to the present $8\times8$ SU(2) lattice study. The baseline and candidate local Metropolis step sizes are varied separately in the main comparison table, Table~\ref{TAB}, and are therefore not listed here.}
\label{TAB3}
\end{sidewaystable}

%------------------------------

%------------------------------
\end{document}